\documentclass[
superscriptaddress,
amsmath,amssymb,
aps,pra,
10pt]{revtex4-2}

\usepackage{bm}

\usepackage{color}
\usepackage{soul}
\usepackage{orcidlink}

\usepackage{stmaryrd}
\usepackage{graphicx}
\usepackage{textcomp}
\usepackage{calrsfs}
\usepackage{yfonts}
\usepackage{physics}
\usepackage[english]{babel}

\usepackage{mathrsfs}
\usepackage{appendix}
\usepackage{bm}
\usepackage{bigints}

\newcommand{\ovn}{\overline{n}}
\newcommand{\B}{\mathrm{B}}
\newcommand{\D}{\mathrm{D}}
\newcommand{\Reff}{R_\mathrm{eff}}
\newcommand{\bul}{\mathrm{bulk}}
\newcommand{\sla}{\mathrm{slab}}
\newcommand{\sph}{\mathrm{sph}}
\newcommand{\cyl}{\mathrm{cyl}}
\newcommand{\uni}{\mathrm{univ}}
\newcommand{\cas}{\mathrm{Cas}}

\newcommand{\cF}{\mathcal{F}}

\newcommand{\cS}{\mathcal{S}}

\begin{document}

\title{Casimir effect with dielectric matter in salted water \\and implications at the cell scale}

\author{Larissa In\'acio \orcidlink{0000-0001-8971-0591}}
\affiliation{Centre of Excellence ENSEMBLE3, Sp. z o. o., Wolczynska Str. 133, 01-919, Warsaw, Poland 
}%

\author{Felipe S. S. Rosa \orcidlink{0000-0001-6187-5992}} 
\affiliation{Instituto de F\'{i}sica, Universidade Federal do Rio de Janeiro, 21941-972 Rio de Janeiro, Rio de Janeiro,  Brazil 
}%

\author{Astrid Lambrecht \orcidlink{0000-0002-5193-1222}}  
\affiliation{%
Forschungszentrum Jülich, 52425 Jülich, Germany
}%
\affiliation{%
RWTH Aachen University, 52062 Aachen, Germany
}
\author{Paulo A. Maia Neto \orcidlink{0000-0001-5287-172X}} 
 \email{pamn@if.ufrj.br}
\affiliation{Instituto de F\'{i}sica, Universidade Federal do Rio de Janeiro, 21941-972 Rio de Janeiro, Rio de Janeiro,  Brazil 
}%

\author{Serge Reynaud \orcidlink{0000-0002-1494-696X}}
\email{serge.reynaud@lkb.upmc.fr}
\affiliation{%
 Laboratoire Kastler Brossel, Sorbonne Université, CNRS, ENS-PSL, Collège de France, 75252 Paris, France  
}%

\date{\today}

\begin{abstract}
The Casimir interaction in salted water contains a universal contribution of electromagnetic fluctuations that makes it of a longer range than previously thought. The universal contribution dominates non universal ones at the distances relevant for actin fibers inside the cell. We discuss universal and non-universal contributions with a model mimicking biological matter. We also show that the universal Casimir effect should have important implications at the cell scale.
\end{abstract}

\maketitle


\section{Introduction}

Quantum physics was born when Planck wrote the first quantum law where he explained the properties of black body radiation~\cite{Planck1900}.
This law gave the mean energy per mode of the electromagnetic field as the product of the energy of a photon $\hbar \omega$ by a mean number $\ovn_\omega$ of thermal photons per mode at frequency $\omega$.
Approximately ten years later, Planck~\cite{Planck1911}
added an extra term $\tfrac12\hbar\omega$ to the mean energy per mode.
This marked the appearance in physics of the zero-point fluctuations which have superseded the now abandoned reasoning of the paper in which they were introduced \cite{Klein1966,Darrigol1988}.

While the original Planck's law fell off the expectation $k_\B T$ by a constant offset ($-\frac{\hbar\omega}{2}$) at high temperatures ($T\to\infty$), the second law including zero-point fluctuations had the correct classical limit  \cite{Einstein1913}. The modern writing of the mean energy per mode makes this property obvious since all terms in the high-temperature expansion go to 0 at $T\to\infty$ but the dominant one
\begin{equation}
 \left( \frac 12 + \ovn_\omega \right) \hbar \omega = \frac
{\hbar \omega}{2} \coth \frac{\hbar\omega}{2k_\B T} \quad
\underset{T\to\infty}{\longrightarrow} \;
k_\B T +O\left(\frac1T\right)~. 
\label{eq:Planckmodern}
\end{equation}
Discussions on zero-point fluctuations were active before their precise nature was better understood as the full development of quantum theory took place  \cite{Milonni1991}.
Vacuum fluctuations, the zero-point fluctuations of electromagnetic fields in empty space, were introduced in old quantum theory \cite{Nernst1916,Kragh2012}.

Nowadays, zero-point fluctuations are considered as a consequence of the non-commutative character of quantum observables \cite{Heisenberg1925,Born1925,Dirac1925,Born1926} (these papers are reproduced and commented on \cite{VanDerWaerden1967book}, where English translations are given for papers written in German). The dispersions of these fluctuations are specified by inequalities deduced from the commutators of observables \cite{Heisenberg1927} (history discussed,  for example, in \cite{Darrigol1992book,Fedak2009,Duncan2023}).
Vacuum fluctuations are now defined by the quantum theory of electromagnetic field \cite{Dirac1927}, with each field mode represented by canonical variables analogous to those of an harmonic oscillator. This theory leads to a full quantum derivation of Einstein's description of absorption, stimulated and spontaneous emission processes \cite{Einstein1917}. Vacuum fluctuations have many important effects in atomic and subatomic physics, such as the Lamb shift and radiative corrections \cite{Power1966ajp,DupontRoc1978l,Itzykson1985book,CohenTannoudji1992book,Milonni2013book}. 

In this paper, we focus the discussions on the mechanical effects due to the radiation pressure of these fluctuations that are Casimir forces~\cite{Casimir1948}, and related  Casimir-Polder and van der Waals forces \cite{Milton2001book,Power2001,Parsegian2005book}.
In his initial work, Casimir studied an idealized problem with two perfectly reflecting plane mirrors in electromagnetic vacuum at zero
temperature. Since then, a large number of papers has generalized the description to make it more realistic in terms of material properties and geometry (a review with many references can be found in \cite{Reynaud2017}). Meanwhile, more and more sophisticated experiments have, in particular, measured precisely the forces between metallic reflecting  surfaces in empty space (reviews and references in \cite{Dalvit2011book,Palasantzas_2015,Woods2016,Stange2021,Milton2022book,Shelden2024}).

We stress at this point that most experiments are performed at room temperature, so that thermal and zero-point fluctuations have to be considered together. The study of Casimir effect in the presence of thermal fluctuations has a long history~\cite{Mehra1967,Brown1969,Schwinger1978}. It is conveniently treated by using the Matsubara summation technique \cite{Matsubara1955}. The latter is directly related to formula 
\eqref{eq:Planckmodern} as Matsubara frequencies are nothing but the poles of the hyperbolic cotangent function, namely \cite{Lambrecht2006}
\begin{equation}
\frac{1}{2} \coth \frac{\hbar\omega}{2k_\B T} =
\sum_{n=-\infty}^{+\infty} \frac{k_\B T}{\hbar\left(\omega-\imath{\xi_n}\right)}  \quad,\quad
\xi_n =n \frac{2\pi k_\B T}{\hbar}  ~. 
\label{eq:matsubara}
\end{equation}
An advantage of the Matsubara technique is that the high-temperature limit of the interaction is given by the term at zero frequency, that is, the pole at $\omega=0$ in formulas (\ref{eq:Planckmodern}-\ref{eq:matsubara}). This contribution has a universal character \cite{Feinberg2001,Canaguier2012} since it does not depend on details of the variation with frequency of the electromagnetic response functions (details discussed below).

For the most common experimental configuration using metallic scatterers in empty space (\cite{Canaguier2012} and references therein), this universal limit can only be met at distances larger than the thermal wavelength in vacuum, where the force is very small and difficult to measure with precision (see the experiments discussed in \cite{Decca2011,Bimonte2021}).
There exists, however, another configuration of great interest in this respect, which involves dielectric matter in electrolytes. In this case, the Casimir force can be both large and dominated by the universal term at not too large distances.
This configuration has been studied in previous theoretical and experimental papers \cite{Ether2015,Neto2019,Pires2021,Schoger2022prl,Schoger2022ijmpa} and will be reviewed in the following. 

This universal Casimir effect in electrolytes is of particular relevance at the interface of physics and biology, as it may have a large impact on the mechanics of biological systems at the cell scale, a point that has been recently emphasized \cite{Spreng2024} and will also be revisited below.

\section{Scattering theory of Casimir interaction in electrolytes}
\label{sec:scattering}

\begin{figure}[h]
    \centering
    \includegraphics[width=0.5\linewidth]{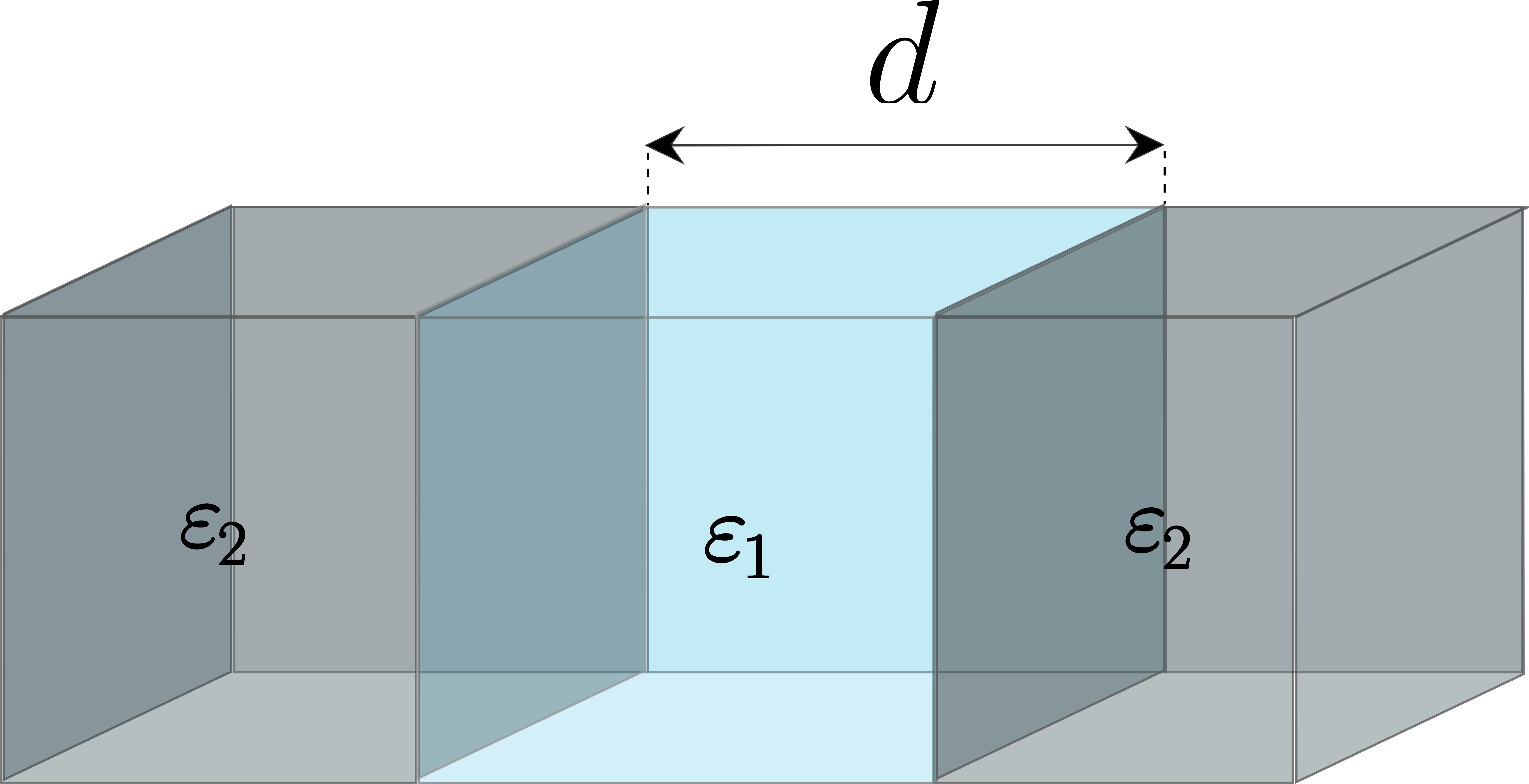}
    \caption{Sketch view of the two-bulks configuration, with $d$ the distance between the two parallel interfaces, $\varepsilon_1$ and $\varepsilon_2$ the dielectric functions for water and immersed matter. }
    \label{fig:twobulks}
\end{figure}

We first describe in this section the scattering theory of the Casimir effect in electrolytes in the simplest geometry of two bulks of matter with parallel plane interfaces, separated by an electrolyte layer \cite{Neto2019}, see Fig.~\ref{fig:twobulks}. 

As we consider the Casimir interaction between objects immersed in an electrolyte, the key ``novelty'' is that 
the dissolved ions give rise to important nonlocal effects. Such nonlocality, also referred to as spatial dispersion, manifests itself in a $k$-dependence of the permittivity in reciprocal space and also in the emergence of a nontrivial tensorial character of the electrolyte permittivity 
\begin{equation}
    \epsilon(\omega) \, \longrightarrow \, \epsilon_{ij}(\omega, {\bf k}) 
    \label{eq:epsilonepsilon}
\end{equation}
The main consequence of the nonlocality of the medium is the support of longitudinal modes, besides the usual transverse electromagnetic fluctuations. The medium allows for 3 independent polarizations: transverse electric (TE), transverse magnetic (TM) and longitudinal ($\ell$), and the extra degree of freedom for the fluctuations has quite important consequences for dispersion interactions \cite{Neto2019}.  

The most economical way of writing the Casimir energy is
by reinterpreting it as a reverberation problem for the electromagnetic fluctuations, and then making a Wick rotation to imaginary frequencies \cite{Lambrecht2006}.  In order to keep things as simple as possible, we assume that the bulk materials are strictly (spatially) local. The Wick rotation produces a Matsubara sum over the frequencies defined in (\ref{eq:matsubara})
\begin{equation}
\begin{aligned}
    \cF_{\cas}=\cF_0+\sum_{n=1}^{\infty} \, \cF_n \,.
   \label{eq:Casimirenergy} 
\end{aligned}
\end{equation}
In the total Casimir free energy $\cF_{\cas}$, we have separated the term $\cF_0$, corresponding to the zero Matsubara frequency, as it  plays the main role in the discussions of this paper. The other terms $\cF_n,$ each corresponding to a nonzero frequency $\xi_n,$ are discussed for the two-bulks geometry in \cite{Neto2019}, and they will be discussed for the two-slabs geometry in \S \ref{sec:nonuniversal} below. 

It can be shown  \cite{Neto2019} that the TM and $\ell$ fluctuations are coupled at the interfaces between the bulks and medium, but, fortunate as it may seem, it turns out that the TE term vanishes in the limit $\xi \rightarrow 0$  and  the TM-$\ell$ contributions actually uncouple from each other \cite{inacio2025casimir}, thus yielding for an area $A$
\begin{equation}
\cF_0 = \underbrace{A \,\frac{ k_{\rm B}T}{2} \!\!
 \bigintsss \! \frac{\mathrm{d}^2k}{(2\pi)^2} \ln \left(1- e^{-2 k d} \right)}_{{-A\,\frac{k_{\rm B}T}{16\pi}  \frac{\zeta (3)}{d^2}}} + \, \underbrace{A\, \frac{ k_{\rm B}T}{2} \!\! \bigintsss \!\frac{\mathrm{d}^2k}{(2\pi)^2} \ln\left(1-  r_\ell^{2} \,e^{-2\sqrt{k^2+{\lambda_\D^{-2}}} d} \right)}_{{\mathcal{F}_{0}}^{(\ell)}} 
\,. \label{eq:TM-L}
\end{equation}
We rewrite this expression by highlighting the first (TM) term that will henceforth be referred to as the \textit{universal} Casimir contribution $ \cF_\uni^\bul$
as it depends only on the temperature and geometry 
\begin{equation}
\cF_0  = \cF_\uni^{\bul} + \cF_0^{(\ell)} \quad,\quad
\cF_\uni^{\bul} = - \frac{A~H}{12\pi\, d^2} \quad,\quad
H = \frac{3\,\zeta (3)}4 \, k_\B T \,.
\label{eq:universal}
\end{equation}
The Hamaker constant $H$ depends only on temperature  ($H\simeq0.9\, k_\B T$ with $\zeta$ denoting the Riemann function and $\zeta(3)\simeq1.202$ the Apery's constant), and it gives $\cF$ through a multiplication by a dimensionless geometrical quantity.
As $\cF_\uni^{\bul}$  is proportional to temperature, the associated entropy is simply deduced to be such that the mechanical energy $\cF_\uni^{\bul} +T\cS_\uni^{\bul}$ vanishes, which means that the universal Casimir interaction is a purely entropic effect. These  properties are still true for the other geometries considered below, with different expressions, naturally. 

The second term in  \eqref{eq:TM-L} is the longitudinal contribution at zero Matsubara frequency, and it is determined by the reflection amplitude $r_\ell$ and the Debye length $\lambda_\D$
\begin{equation}
   r_\ell=\frac{\epsilon_{b}(0)\sqrt{k^2 + \lambda_\D^{-2}} - \epsilon_2(0) k}{\epsilon_{b}(0)\sqrt{k^2 + \lambda_\D^{-2}} +  \epsilon_2(0) k} \hspace{20pt} , \hspace{20pt} \lambda_\D = \sqrt{\frac{\epsilon_{b}(0) k_\B T}{n^2 m}} \, ,
   \label{eq:Debyelength}
\end{equation} 
where $n$ is the ionic concentration in the electrolyte, $m$ is the ion mass, and $ \epsilon_b$ is the dielectric permittivity of pure water (i.e., the permittivity of water without the ionic contribution).

In the long distance limit, the longitudinal part $\cF_0^{(\ell)}$ is exponentially suppressed by the Debye screening with typical length $\lambda_\D$ 
\begin{equation}
    \label{eq:expo}
\cF_0^{(\ell)}\,\propto\, e^{-2d/\lambda_\D} \, .
\end{equation}
This is the first major takeaway of this paper:  there is indeed a screening of the contribution $\cF_0^{(\ell)}$ of longitudinal modes across the electrolyte, but the universal contribution $\cF_\uni^{\bul}$ is \emph{unscreened} as it arises from the transverse electromagnetic fluctuations.
This property was first demonstrated in the framework of a macroscopic description of the dielectric response of salted water  \cite{Neto2019} and it has been confirmed since then by the molecular dynamics simulations presented in \cite{Spreng2024}. 
This has far-reaching consequences for physical and even biological systems, which we will discuss in the following. 

 An important remark has to be made clear at this point. The discussions just presented on transverse and longitudinal fluctuations treat fields in the vicinity of zero frequency rather than at zero frequency. In technical terms, what has been calculated is given by a residue associated with the pole at $\omega=0$ of the cotangent function in \eqref{eq:matsubara}. The physical origin of the effect is simply the divergence as $\frac{k_\B T}{\hbar\omega}$ of the number of photons in a field mode with frequency close to $\omega=0$, leaving in the end a finite contribution to the free energy.

\section{Experimental evidence with optical tweezers}
\label{sec:experiment}

Optical tweezers~\cite{Ashkin2007} are an ideal tool to measure interaction forces in aqueous media, and several applications in biology have been developed~\cite{Nussenzveig2015, Bustamante2021} over the years. Since the trap stiffness is proportional to the beam laser power,  the system can be tuned to match the required order of magnitude. Standard single-beam traps are most effective for particles whose refractive indexes are slightly higher than that of the surrounding medium. 
Conveniently, this is also the condition that minimizes the non-universal contributions $\cF_{n\geq1}$ which depend on the reflection (or more generally scattering) amplitudes at the interface between the material and the host medium. 

\begin{figure}[h]
    \centering
    \includegraphics[width=0.6\linewidth]{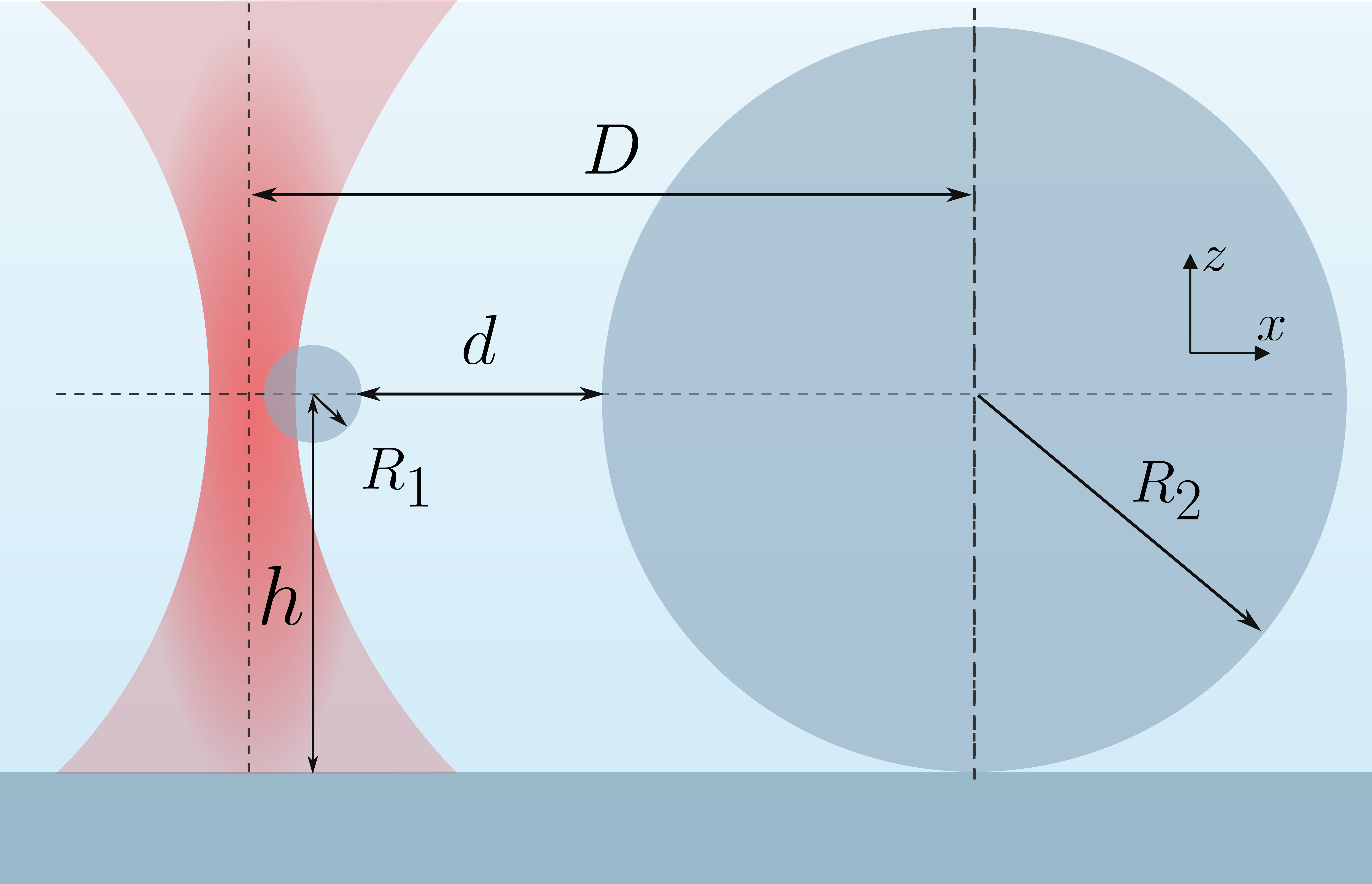}
    \caption{Principle of the experiment reported in Ref.~\cite{Pires2021}: a silica microsphere (radius $R_1$) is held by a tightly focused laser beam close to a larger silica microsphere (radius $R_2$) which is attached to the glass slide at the bottom of the sample chamber. The distance $D$ between the larger microsphere and the laser axis is controlled by using a piezoelectric nano-positioning system. The Brownian fluctuations of the trapped microsphere are measured for different values of $D$. The distance of closest approach of the two spheres is denoted $d$ and the height of the small sphere above the glass slide $h$. }
    \label{fig:experimentscheme}
\end{figure}

The universal Casimir interaction 
between two silica microspheres immersed in salted water
has been measured with optical tweezers~\cite{Pires2021}.   The experimental scheme is sketched in Fig.~\ref{fig:experimentscheme}. 
Brownian fluctuations of a small silica microsphere (radius $R_1=2.4\,\mu{\rm m}$) held by optical tweezers were measured along the $x$ and $y$ directions transverse to the laser beam.
The interaction energy was then observed through the modification of the fluctuations along the $x$-axis connecting the sphere centers, 
as the larger microsphere (radius $R_2=12\,\mu{\rm m}$), adhered to the glass slide at the bottom of the sample, was approached
by employing a piezoelectric nano-positioning system. The sample was also displaced vertically to align the sphere centers so as to have $h=R_2,$ where $h$ is the height of the small microsphere with respect to the glass slide.
While fluctuations along $x$ captured the interaction signal, fluctuations
along  $y$ were not modified, showing the absence of optical binding or of any other perturbation of the optical force in this configuration with near index matching. 
 Since the trapping beam was circularly polarized, the Brownian fluctuations along $y$ accumulated over several experimental runs were employed to infer the optical potential relevant for the interaction direction $x.$
 As an additional check, the resulting optical potential was shown to agree with the Mie-Debye theory of optical tweezers~\cite{Dutra2007,Dutra2014} as well as with standard Stokes calibration~\cite{Sarshar2014}, with the latter implemented in the absence of the larger microsphere. 

The interaction energy for distances above $100\,{\rm nm}$
was then obtained by subtracting the optical potential from the total energy as determined by the Brownian dynamics along the $x-$direction. For moderate salt concentrations, the interaction was dominated by the 
electrostatic double layer interaction.  
In order to suppress such interaction, a second experiment reported in Ref.~\cite{Pires2021} employed a high salt concentration (strong screening with $\lambda_\D=0.65\,{\rm nm}$).  In this second case, 
the interaction was completely dominated by the universal Casimir contribution discussed in the previous section.

\begin{figure}[h]
    \centering
    \includegraphics[width=0.64\linewidth]{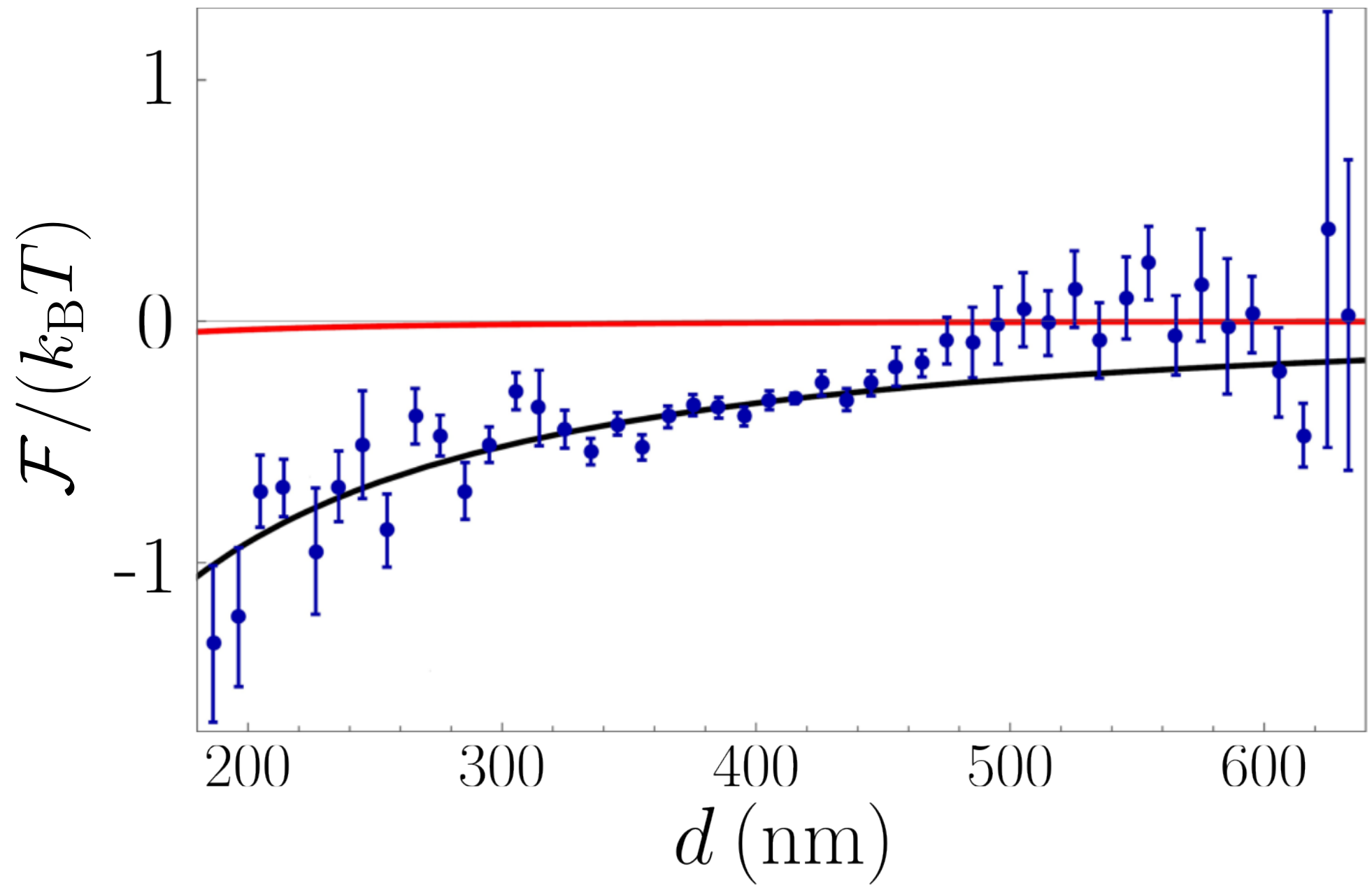}
    \caption{Variation of the Casimir interaction energy (in units of the thermal energy $k_\B T$) with the distance $d$ between two dielectric microspheres in salted water for the setup depicted in Fig.~\ref{fig:experimentscheme}. Experimental points are shown with their error bars (blue). The old approach relying on overlooking the universal contribution (red) leads to much too small values to be compatible with the experimental data. New theory (black), with the universal contribution included, agrees fairly well with experiments (with no fitting). Adapted 
 from Pires et al.~\cite{Pires2021} under the terms of the Creative Commons Attribution 4.0 International license.   }
    \label{fig:experimentalplot}
\end{figure}

Figure~\ref{fig:experimentalplot} shows the resulting Casimir energy in units of $k_\B T$ versus distance $d$ of closest approach (see Fig.~\ref{fig:experimentscheme}). The red curve represents the non-universal contribution accounting for the positive Matsubara frequencies $\xi_{n\geq1}$.  The calculation is based on Mie scattering by dielectric spheres in water~\cite{Spreng2020}.  The red curve thus represents the 
generalization  for the geometry of two spheres of the
standard theoretical prediction for the Casimir interaction in salted water~\cite{Parsegian2005book}, which is usually considered within the parallel planar interfaces setup or within the corresponding  proximity force approximation. 
The longitudinal contribution $\cF_0^{(\ell)}$ is not included as it 
is negligible at both moderate and high salt concentrations employed in Ref.~\cite{Pires2021}, due to the fact that the distances are much larger than the corresponding Debye screening lengths. Indeed, 
$\cF_0^{(\ell)}$ is
given by the second term in the right-hand-side of Eq.~(\ref{eq:TM-L}) for parallel planar interfaces and calculated for two spheres in Ref.~\cite{Nunes2021}. In both cases, 
$\cF_0^{(\ell)}$ becomes exponentially small 
 at distances larger than $\lambda_{\rm D}.$

The black curve in  Fig.~\ref{fig:experimentalplot} stands for the total Casimir interaction, including the universal contribution arising from TM modes in the zero frequency limit. 
Like the non-universal contribution, it is calculated exactly for the geometry of two dielectric spheres in salted water as described in Refs.~\cite{Schoger2022prl,Schoger2022ijmpa} (see also Sec.~\ref{sec:geometry} below).
The comparison between the black and red curves indicates that the universal contribution dominates the Casimir interaction for distances above $200\,{\rm nm}.$ More importantly, the unscreened, universal Casimir contribution provides an excellent description of the data, with no fitting parameter, in contrast with the standard approach that overlooks the contribution of TM modes and is excluded by the experimental data shown in Fig.~\ref{fig:experimentalplot}.

\section{Universal Casimir interaction in the two-spheres or two-cylinders geometries \label{sec:geometry}}

The universal Casimir interaction that is the high-temperature limit of the expression in the case of very efficient Debye screening has been calculated in the two-spheres \cite{Schoger2022prl,Schoger2022ijmpa} and two-cylinders \cite{Spreng2024} geometries. As in the two-bulks geometry considered previously, it arises from the term $n=0$ in the Matsubara sum appearing due to transverse electromagnetic fluctuations, and it does not depend on details of the frequency dependence of dielectric functions. 

\begin{figure}[h]
    \centering
    \includegraphics[width=0.9\linewidth]{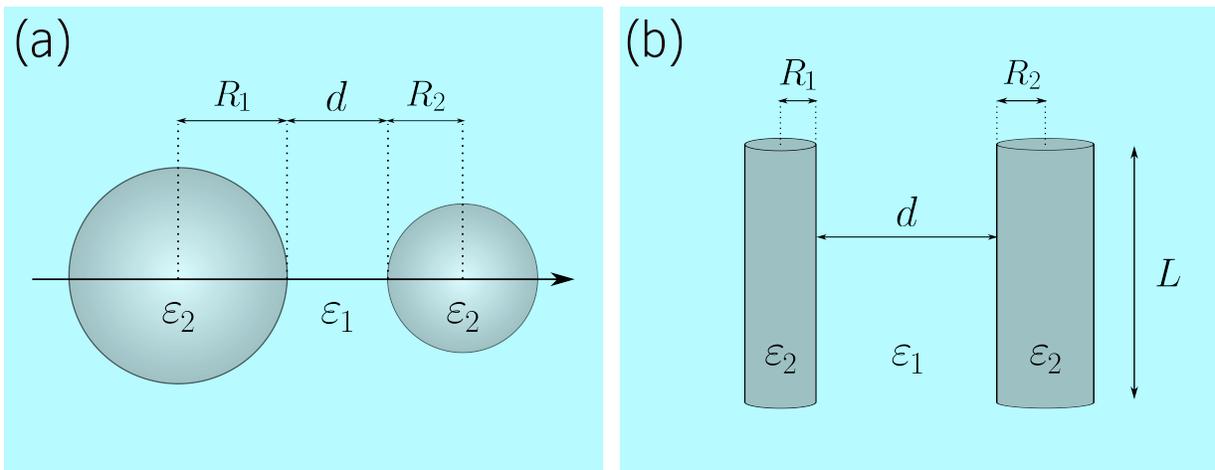}
    \caption{ Geometry for two spheres (a) or two cylinders (b) immersed in salted water. $R_1$ and $R_2$ are the radii of spheres or cylinders, and $d$ the distance of closest approach ($\varepsilon_1$ and $\varepsilon_2$ defined as for Fig.\ref{fig:twobulks}). }
    \label{fig:scheme}
\end{figure}

 The associated free energy is the product of the thermal energy scale $k_\B T$ by a dimensionless function $f$ which now depends in a non trivial manner on the geometrical parameters. In the two-spheres geometry, we have  
\begin{equation}
\cF_\uni^\sph = - k_\B T \, f\left(d,R_1,R_2\right)  ~. \label{eq:universalsphere}
\end{equation}
The Casimir interaction is captured in the function $f$ which has been calculated for two dielectric spheres in salted water \cite{Schoger2022prl,Schoger2022ijmpa}. 
It does not depend on temperature (the effect is purely entropic) and it is a dimensionless function of the two ratios which can be formed with the three length parameters $d$ (note that it was denoted $L$ in \cite{Schoger2022prl,Schoger2022ijmpa}), $R_1$ and $R_2$ (radii of the two spheres; see Fig.\ref{fig:scheme}a).

The function $f$ is calculated within the scattering theory of the Casimir effect \cite{Lambrecht2006}. It is a sum over all field modes of scattering operators involving an arbitrary number of round-trips in the cavity formed by the two objects.
In the so-called dipolar limit where the spheres have small sizes in comparison with the distance, the contributions of large numbers of round-trips are negligible. In this small size approximation (SSA) or equivalently long distance approximation, $f$ is proportional to the two volumes $R_1^3$ and $R_2^3$, and inversely proportional to the sixth power of distance~\cite{Schoger2022prl,Schoger2022ijmpa}
\begin{equation}
f_\mathrm{SSA} =\frac{3 R_1^3~R_2^3}{4d^6}\ \;.\label{eq:SSAspheres}
\end{equation}
In the opposite limit of a short distance between the spheres, the exact result involves large numbers of round-trips, and it tends to an expression related to the two-bulks geometry through the proximity force approximation~\cite{Spreng2018}. In this PFA limit, $f$ is proportional to the effective radius $\Reff$ defined from the two radii $R_1$ and $R_2$, and inversely proportional to the distance 
\begin{equation}
 f_\mathrm{PFA} =\zeta(3)\frac{\Reff}{8d}  \quad,\quad
\Reff =\frac{R_1R_2}{R_1+R_2} \;.
 \label{eq:PFAspheres}
\end{equation}

The function $f$ has a universal expression which does not depend on the details of the variation with frequency of the permittivity functions of either the dielectric matter in the spheres or of salted water constituting the immersion medium. The result is conveniently conveyed - just like in Fig.~1(b) of \cite{Schoger2022prl} - as a function $f_u\left(x\right)$ of two dimensionless parameters 
\begin{equation}
x =\frac{d}{\Reff}  \quad,\quad
u =\frac{R_1\,R_2}{\left(R_1+R_2\right)^2} ~,
 \label{eq:dimensionlessparameters}
\end{equation}
where $x$ is an aspect ratio for the distance $d$, and $u$ an aspect ratio for the radii. The fact that $f$ depends only on $x$ and $u$ is a remarkable scale-invariance  property meaning that $f$ is unchanged when the three geometrical dimensions $d,R_1,R_2$ are multiplied by a common factor. 

A more subtle property appears when $f$ is written (see Fig.~2 in \cite{Schoger2022prl}) in terms of another parameter
\begin{equation}
y=\frac{\left(d+R_1+R_2\right)^2+R_1^2+R_2^2}{2R_1\,R_2} 
= 1+x\left(1+\frac{ux}{2}\right)~.
 \label{conformalparameter}
\end{equation}
The value of $f$ thus appears to depend mainly on $y$, which reveals an approximate conformal invariance of geometrical dependence of the free energy \cite{Schoger2022prl,Schoger2022ijmpa} .
This property has been used in \cite{Schoger2022ijmpa} to give a simple formula which can be applied to a wide range of geometrical parameters. This formula is sufficient for the salinity of biological media and at ambient temperature. 
The obtained free energy $\cF_\uni^\sph$ is larger than $k_\B T$ ($f\geq1$) only for spheres very close to each other, that is the geometry in which the universal Casimir interaction has been experimentally measured (see \S\ref{sec:experiment}).

\begin{figure}[h]
    \centering
        \includegraphics[width=0.49\textwidth]{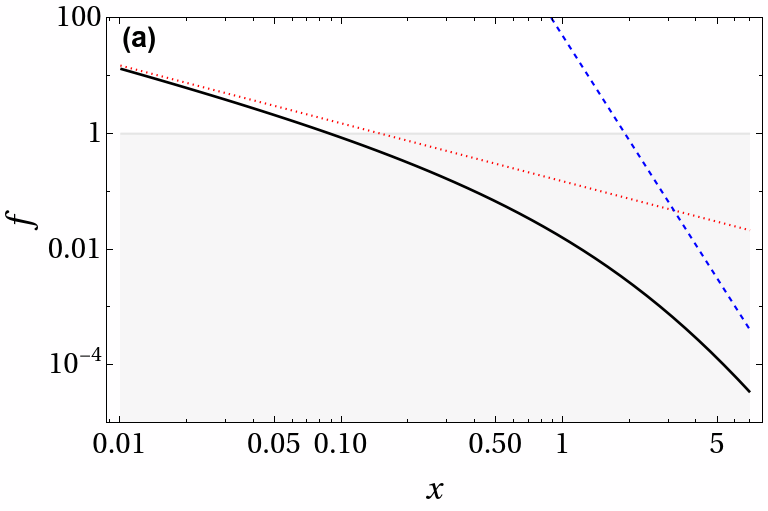}
        \includegraphics[width=0.49\textwidth]{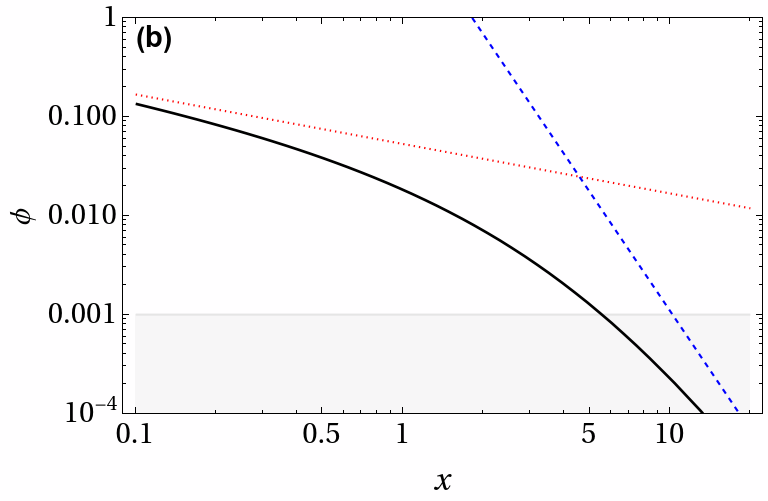}
 \caption{Universal functions (solid black curves) showing the variations of the universal free energy versus $x=d/\Reff$:   
 (a) function $f$ calculated for two spheres with equal radii; 
 (b) function $\phi$ calculated for two cylinders with equal radii. 
 In both plots, red dotted and blue dashed lines represent the  proximity force and small size approximations giving the asymptotic behaviors at $x\ll1$ and $x\gg1$ respectively.}
    \label{fig:phiversusx}
\end{figure}

In order to compare the two-spheres case and the two-cylinders one discussed below, we consider the geometry of two spheres with equal radii $R_1=R_2=R$ and draw in Fig.~\ref{fig:phiversusx}a the corresponding function $f$ versus the geometric parameter $x$ defined in \eqref{eq:dimensionlessparameters}.
The grayed out band in the figure shows where $f<1$, which is also where the binding energy $\vert\cF\vert$ is smaller than the Brownian motion energy scale. The figure shows that the universal Casimir effect dominates the Brownian energy $k_\B T$ for distances 
$d<0.09\,R_{\rm eff}$.
Note that, for the geometry with different radii employed in the optical tweezers experiment of \S\ref{sec:experiment}, , the order of magnitude for this border $|\cF_\uni^\sph| \simeq k_{\rm B}T$ is at $d\approx 0.2\,\mu{\rm m},$ which is indeed the scale of distances probed in Ref.~\cite{Pires2021}.

We now come to the discussion of the configuration with two dielectric cylinders in salted water. This configuration makes the interaction larger than in the two-spheres case as it scales as the length $L$ of the cylinders. The geometry is shown in Fig. \ref{fig:scheme}b with two cylinders with radii $R_1,R_2$ at a distance $d$ of closest approach.
In this geometry, the universal Casimir free energy is the product of the thermal energy $k_\B T$ by two dimensionless factors $\frac Ld$ and $\phi$ 
\begin{equation}
\cF_\uni^{\cyl} = - k_\B T \, \frac{L}{d} \, \phi\left(d,R_1,R_2\right)  ~. \label{eq:universalcylinder}
\end{equation}
The physics of the universal interaction is captured in the function $\phi$ which has been calculated in \cite{Spreng2024}.  
The remarks written after equation \eqref{eq:universalsphere} remain valid here as $\phi$ does not depend on temperature and is a dimensionless function of two ratios of the three geometrical quantities $d, R_1$ and $R_2$. 

In order to discuss the implications of these results for biological systems in section \ref{sec:implications}, we depict in Fig.~\ref{fig:phiversusx}b  the function $\phi$ for two cylinders with equal radii $R_1=R_2=R$.
The function $\phi$ is plotted versus the geometric parameter $x$ defined as in \eqref{eq:dimensionlessparameters}, but for cylinders.
The grayed out band in the figure shows where $\phi<0.001$, that is where the binding energy $\vert\cF\vert$ is smaller than the Brownian motion energy scale for $L=1000~d$ (see the discussions to be presented in section \ref{sec:implications}). The take-home message from the figure is that the universal Casimir free energy dominates the Brownian motion energy $k_\B T$ imposed by the surrounding water over a broad range of parameters, including in particular distances $d$ of the order of $R_{\rm eff}$. 
The striking difference between the two-spheres case (Fig.~\ref{fig:phiversusx}a) and the two-cylinders one (Fig.~\ref{fig:phiversusx}b) is simply due to the presence of the large factor $\frac Ld$ in formula \eqref{eq:universalcylinder} which makes $\vert \cF_\uni^{\cyl}\vert$  larger than $k_\B T$ in a wider range of parameters. 

Both functions in Fig.~\ref{fig:phiversusx} agree with the proximity force approximation (red dotted curve) at short separations and with the small size approximation (blue dashed line) at long distances. For completeness, we give the PFA and SSA expressions for the case of two cylinders 
\begin{equation}
\phi_\mathrm{PFA} = \frac{\zeta(3)}{32}\,\sqrt{\frac{2\Reff}{d}} \quad,\quad
\phi_\mathrm{SSA} = \frac{891\pi}{4096}\,\frac{R_1^2~R_2^2}{d^4}
~. \label{eq:PFAcylinders}
\end{equation}
These expressions have to be compared to eqs.~\eqref{eq:PFAspheres} and \eqref{eq:SSAspheres} respectively. In particular the changes of power laws have clear meanings. 
For the PFA,  $\phi$ is proportional to $\sqrt{{\Reff/d}}$ because there is only one curvature radius in a direction orthogonal to the distance for cylinders (there are two of them for spheres and $f$ is proportional to ${{\Reff/d}}$). For the SSA, the extensive quantity in the perturbative limit is the area for cylinders rather than the volume for spheres.
It follows that $\phi$ is proportional to the two areas $R_1^2$ and $R_2^2$, and inversely proportional to the fourth power of distance.

\section{Non-universal contributions with dielectric matter in salted water}
\label{sec:nonuniversal}

The universal Casimir interaction was calculated in \cite{Spreng2024} for two dielectric cylinders in salted water, but the non universal contributions have never been analyzed in this geometry. In order to fill this gap, we need a robust model for the dielectric response of matter, as non universal contributions depend on this response. In the absence of an exact solution, we have also to produce a reasonably accurate estimate for the interaction between two cylinders. These requirements are tackled in the present section.

\begin{figure}[th]
\begin{center}
\includegraphics[width=0.6\linewidth]{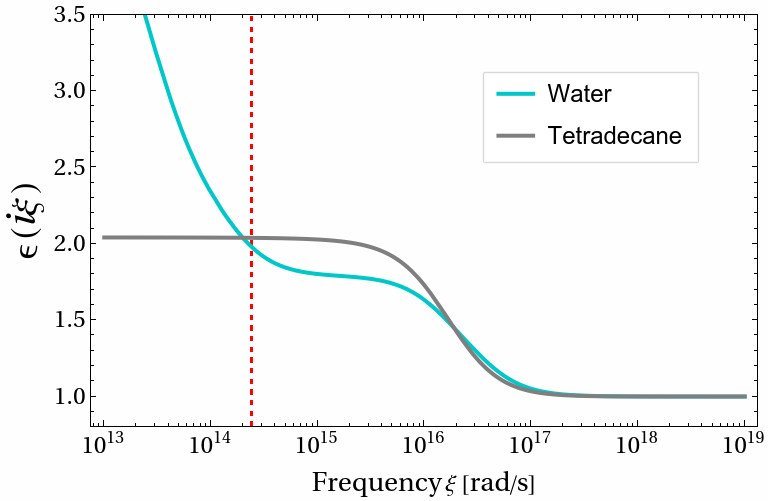}
\end{center}  \caption{Dielectric functions of water (blue) and tetradecane (gray) versus imaginary frequency $\xi$. The vertical dashed (red) line represents the first nonzero Matsubara frequency $\xi_1=2\pi k_\B T/\hbar$ at ambient temperature, where the dielectric functions of water and tetradecane are nearly identical. }
 \label{fig:dielectricfunctions}
\end{figure}

For describing dielectric properties of organic matter, we follow~\cite{Parsegian1970,Parsegian1971} by considering tetradecane as a proxy for these properties.
Its dielectric function is given by a Lorentz model, with parameters to be found in~\cite{Parsegian1981}. In addition, we model salted water as in~\cite{Neto2019,inacio2025casimir}, with the contribution from the ions depending on their concentration. As we mentioned above, this contribution is critical for the universal contribution at the zero frequency, but it plays a negligible role at nonzero Matsubara frequencies, which are too large (at room temperature) to ``feel'' the effect of ionic conduction (when compared to electrons, ions have a much larger mass and their number per unit volume is much smaller). The tetradecane function is represented in Fig.~\ref{fig:dielectricfunctions} versus imaginary frequency $\omega=\imath\xi$ (i.e., after a Wick rotation, as is common for studies of the Casimir effect \cite{Reynaud2017}), alongside the pure water function (i.e., disregarding the ionic contribution) taken from the model in~\cite{vanZwol2010}. The vertical dashed line on the figure shows the position of the first nonzero Matsubara frequency $\xi_1$ at room temperature. The figure indicates a nearly perfect index matching of biological matter and water at this frequency, which is at the root of the smallness of the non universal contributions to the Casimir interaction. Although the contribution of ions is negligible at this frequency, the index contrast grows sharply for lower frequencies, in particular as the effective response diverges for salted water. This Drude-like divergence caused by the dissolved ions is the reason why the universal contribution ends up not depending on the details of the properties of the material immersed in salted water.

In order to get reliable estimations for the non universal contributions to interaction between cylinders, we use the fact that these contributions are expected to be small due to the index-matching at most non null Matsubara frequencies and, therefore, that they should be approximately represented by a pairwise summation technique \cite{Bitbol2013}. We stress that such an approximation cannot give an accurate description of the universal contribution as perturbation theory is not at all adapted to this case. Guided by the pairwise summation, we consider that the order of magnitude of the ratio between universal and non universal contributions arising from two cylinders of radii $R$ can be roughly estimated from the same calculation in a setup of two slabs of thickness $w\equiv 2R$ separated by a gap $d$ of salted water. 

\begin{figure}[h]
    \centering
    \includegraphics[width=0.5\linewidth]{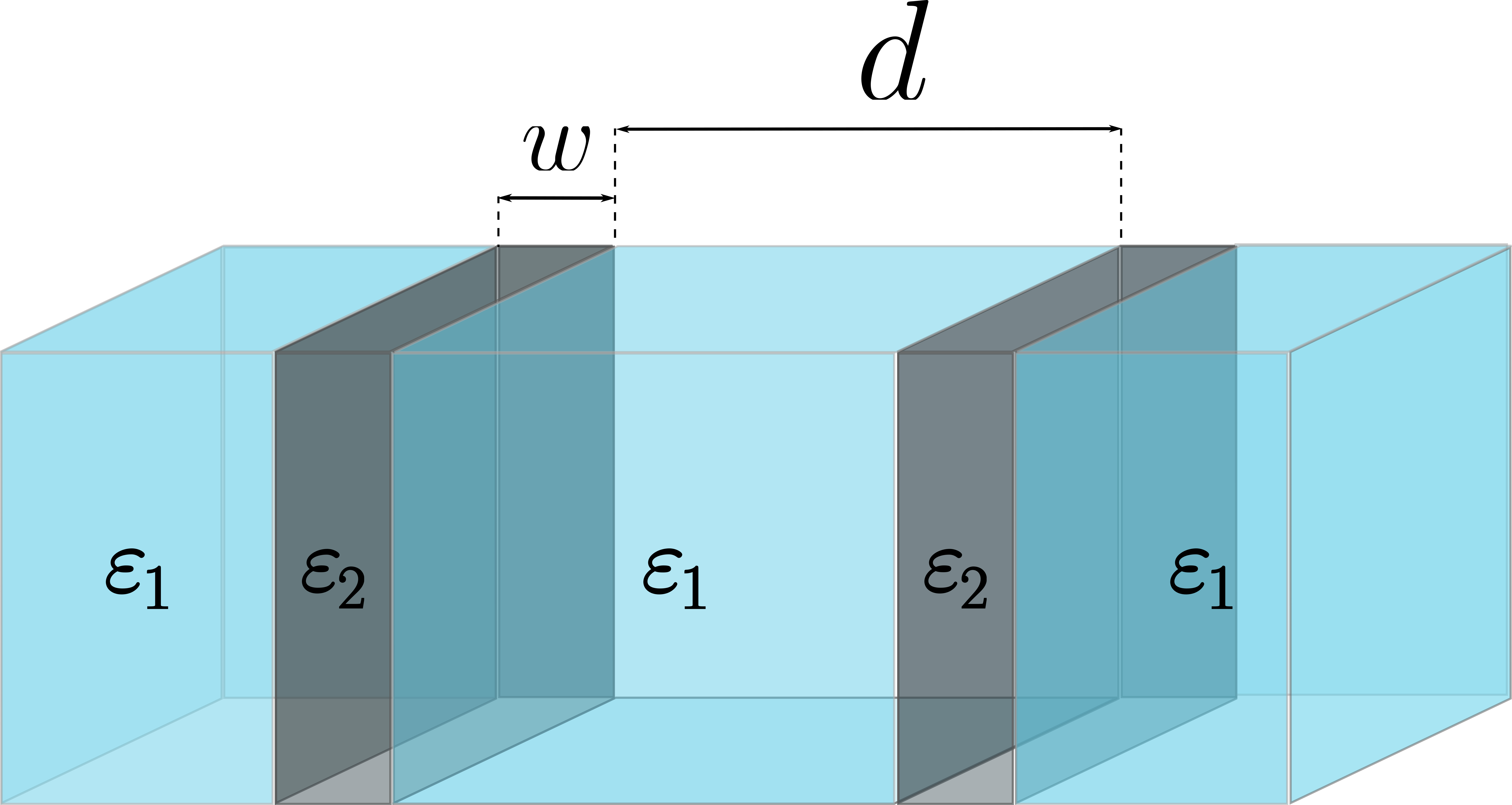}
    \caption{Sketch view of the two-slabs configurations, with $d$ the distance between the two parallel interfaces and $w$ the width of the slabs  ($\varepsilon_1$ and $\varepsilon_2$ defined as for Fig.\ref{fig:twobulks}).}
    \label{fig:slabs}
\end{figure}
Rewriting expression (\ref{eq:Casimirenergy}) for slabs, we get
\begin{equation}
    \cF_{\cas}^{\sla} = \cF_0^{\sla} + 
    \sum_{n=1}^{\infty} \, \cF_n^{\sla} \, .
    \label{eq:CasimirSlabs}
\end{equation}
where the nonzero Matsubara terms are given by
\begin{equation}
    \cF_n^{\sla} = 
A \, k_{\rm B}T\, 
\sum_{\sigma} \bigintsss \frac{\mathrm{d}^2k}{(2\pi)^2} \ln \left(1- \tilde r^{(\sigma)}(i\xi_n)^2 \,  e^{-2\kappa_1 d} \right)\quad,\quad  n\neq 0 \,
\label{eq:nonzeroMatsubara}
\end{equation}
with $\sigma$ standing for either TE or TM polarization of a given electromagnetic mode. The reflection coefficients of a slab are given by ~\cite{Pirozhenko2008} 
\begin{equation}
 \tilde r^{(\sigma)}(\imath\xi_n) = 
 \frac{r^{(\sigma)}(\imath\xi_n) \left(1-\exp(-2\kappa_2\,w)\right)}{1-r^{(\sigma)}(\imath\xi_n)^2\exp(-2\kappa_2\,w)}\quad, \quad 
 \kappa_i = \sqrt{\epsilon_i(\imath\xi_n) \,\xi_n^2/c^2 + k^2}  \,.
 \label{eq:RefCoefSlab}
\end{equation}
where $r^{(\sigma)}(i\xi_n) $ are the Fresnel coefficients for a plane wave going from the electrolyte to the slab
\begin{equation}
r^{(\rm TE)}(\imath\xi_n)=\frac{\kappa_1-\kappa_2}{\kappa_1+\kappa_2} \quad, \quad \quad 
r^{(\rm TM)}(\imath\xi_n)=\frac{\varepsilon_2\kappa_1-\varepsilon_1\kappa_2}{\varepsilon_2\kappa_1+\varepsilon_1\kappa_2}  \,.
\label{eq:RefCoefFresnel}
\end{equation}
The zero-frequency Matsubara term $\cF_0^{\sla}$ for the slabs is analogous to expression (\ref{eq:TM-L}). Given that we are interested in distances $d$ largely beyond the Debye length, the longitudinal part is negligible 
\begin{equation}
    \cF_0^{\sla} \simeq \cF_{\uni} ^{\sla}\, .
\end{equation}
As shown in the Appendix of \cite{inacio2025casimir}, the universal contribution for two slabs is the same as in the case of two half-spaces discussed in 
\S \ref{sec:scattering}: $\cF_{\uni} ^{\sla}=\cF_{\uni} ^{\rm bulk}.$ 

\begin{figure}[htbp]
\begin{center}
\includegraphics[scale=0.5]{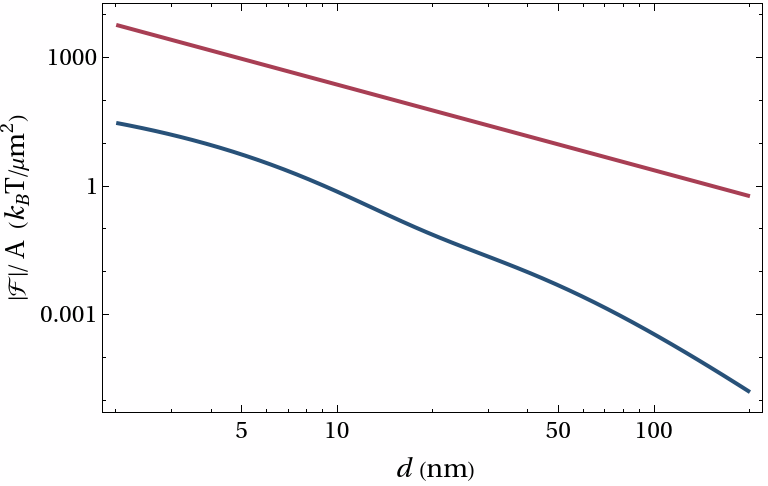}
\end{center}  \caption{Universal (red higher line) and non-universal (blue lower line) contributions to the Casimir binding energy, for two slabs of tetradecane of thickness $w=6\,{\rm nm}$ separated by a salted water gap $d$. The binding energy $\vert\cF\vert$ per unit area are represented in units of thermal energy $k_\B T$ per $\mu {\rm m}^2$. }
 \label{fig:bindingenergy}
\end{figure}

The variations of the 
universal contribution for the binding energy $\vert\cF_\uni^{\sla}\vert$ and the non-universal part $\vert\sum_{n\geq1} \cF_n^{\sla}\vert$ versus  distance $d$ are depicted separately 
in Fig. \ref{fig:bindingenergy},  for identical slabs of thickness $w=6$~nm (this number is motivated by discussions in \S\ref{sec:implications}). The non-universal contributions have been calculated with the dielectric model for tetradecane represented  in Fig.~\ref{fig:dielectricfunctions}. Figure \ref{fig:bindingenergy} clearly shows that the non-universal contributions are much smaller than the universal one, which is a consequence of the index-matching at nonzero Matsubara frequencies between water and tetradecane. We have checked that the ratio between the two contributions is always above a factor 100.

\section{Implications at the cell scale}
\label{sec:implications}

We now use the results presented above to show how the universal Casimir interaction may play an important role in biomechanics at the cell scale. The question was thoroughly discussed in \cite{Spreng2024} where many references can be found. Here we only recall a few critical points. 

Actin filaments play critical roles in the mechanical functions of the cytoskeleton as well as in many intracellular processes, by actively generating forces with the help of motor proteins~\cite{Salbreux12,Murrell15,Burla19}. They form bundles, with filaments cross-linked by specific proteins into parallel arrays. It has been shown that bundles of parallel filaments may form \textit{in vitro} in the absence of cross-linkers~\cite{Tang96,Deshpande12}. Other examples of self-assembled filaments are discussed in  \cite{Spreng2024}, which play key roles in or beyond the cytoskeleton.
The universal Casimir interaction considered above is relevant at the observed dimensions for bundles of actin filaments, so that they should have important implications in the self-assembly and cohesion of bundles of filaments at the cell scale. 

\begin{figure}[h]
\centering
\includegraphics[width=0.6\linewidth]{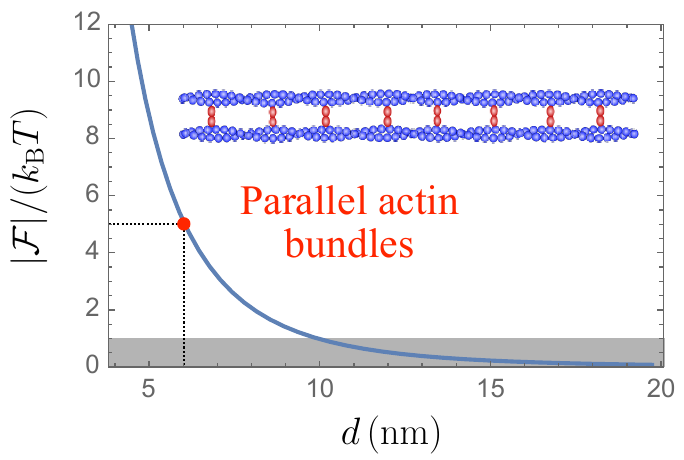}
\caption{Casimir binding free energy between actin
 filaments having each a radius $3\,{\rm nm}$  and a length $15\,\mu$m. The binding free energy $\vert\cF\vert$  is drawn versus the distance $d$ (nm) between  two parallel actin filaments and measured in units of $k_\B T$. In the gray-shaded zone $\vert\cF\vert$ is dominated by the thermal energy $k_\B T$. Energies above this zone are  relevant for biomechanics in the cell, and they include the physiological distance ($6\,{\rm nm}$) indicated by the red marker. Inset: schematic of two actin filaments (blue) in a two-filaments  bundle, with cross-linkers shown in red. Adapted from Spreng et al.~\cite{Spreng2024} under the terms of the Creative Commons Attribution 4.0 International license.}
 \label{fig:actinbundle}
\end{figure}

Estimation of the interaction for the relevant case of parallel actin filaments is done by modeling filaments as cylinders and using the physical dimensions known for  parallel actin filaments~\cite{Volkmann2001}, each having a radius $R\simeq3$~nm and a length $L\simeq15~\mu$m, which form bundles with typical distance of closest approach $d\simeq6$~nm between filaments (geometry sketched as an inset in Fig.~\ref{fig:actinbundle} with filaments shown in blue and cross-linkers in red). 
Such distance is about 10 times larger than the characteristic Debye screening length $\lambda_\D$ in biologically relevant solutions. The electrostatic interaction as well as the longitudinal contribution to the Casimir interaction (see \S.2) are then negligible as they are efficiently screened.
The parameters given above correspond to a large value for the ratio $L/d\simeq2500$, which is a key reason for which the correspondent Casimir interaction is so significant.
The binding energy $\vert\cF\vert$ expressed in units of $k_\B T$
is shown versus the separation distance $d$ in Fig.~\ref{fig:actinbundle}. 
The relevant distance $d\simeq6$~nm is emphasized as the red point on the plot, where the binding energy is of the order of 5 $k_\B T$. In the gray-shaded zone, $\vert\cF\vert$ is dominated by the Brownian motion energy $k_\B T$.

We stress at this point that such a magnitude
had to be expected for a physical interaction having relevance for biomechanics at the cell scale. The interaction has to be larger than $k_\B T$ in order to dominate Brownian agitation while it should not be too large to be compatible with a mechanical activity provided by molecular motors. This was underlined by Moysés Nussenzveig in his remarkable paper~\cite{Nussenzveig2015} commenting on the links between physics and biology, and discussing in particular the pioneering works of Bohr \cite{Bohr1933} and Schr\"odinger \cite{Schrodinger1944}. 

We also emphasize that this magnitude has been obtained without any \emph{ad hoc} fine-tuning of the parameters. The energy constants are given by the theory of the universal Casimir interaction reviewed in this paper, and they do not depend on any specific detail for the biological matter involved. Hence the binding energy depends only on geometrical parameters known for bundles of actin filaments~\cite{Volkmann2001}. Note also that the numbers presented for actin (red marker shown on  Fig.~\ref{fig:actinbundle}) correspond to $x=4$ in Fig.~5b, where the PFA expression overestimates the correct value of $\phi$ by a factor 13. This implies that the calculation presented in \cite{Spreng2024} for cylinders was mandatory for obtaining a reliable estimation for bundles of actin filaments.

Another important point is clear after the discussions in the present paper. We have given an estimation of the non universal Casimir interaction in \S \ref{sec:nonuniversal}, and shown that it is more than a hundred times smaller than the universal one. 
For the relevant distance $d = 6$nm between actin filaments,
the non universal contribution is then smaller than $0.05\,k_\B T$. 
It is thus clear that the non universal interaction is not only much smaller than the universal one, but also much too small to have any relevance in a medium dominated by Brownian agitation. It is only the presence of the universal contribution which makes the Casimir binding energy important for biomechanics at the cell scale. 

As a concluding remark, we want to stress that physics is extremely complex at the cell scale, with a large number of different processes involved in the mechanics and functionalities of the cell. Detailed discussions of this very important fact are presented in the concluding part of \cite{Spreng2024}. Our modest aim in this review has only been to show that the universal Casimir interaction due to transverse electromagnetic fluctuations has to be considered as a significant contribution to this rich physics as it produces an attractive interaction between biological filaments in the cell with a typical energy of a few Brownian thermal scales $k_\B T$.

\section*{Acknowledgments}

This paper is dedicated to the memory of our dear and esteemed colleague, H. Moysés Nussenzveig, to whom we are indebted for many inspiring discussions.

The authors warmly thank the colleagues who have participated in the development of this project over the years, in particular 
D.S. Ether, L.B. Pires, S. Umrath, D. Martinez, Y. Ayala, B. Pontes, G.R.S. Araújo, S. Frases, G.-L. Ingold, N.B. Viana, 
R.O. Nunes, A.B. Moraes, A. Canaguier-Durand,
R. Guérout, B. Spreng, M.J.B. Moura, R.S. Decca, R.S. Dutra,
M. Borges, T. Schoger, H. Berthoumieux, A.-F. Bitbol.

This work was supported by CAPES-COFECUB collaboration project n° Ph1030/24,  
Conselho Nacional de Desenvolvimento Cient\'{\i}fico e Tecnol\'ogico (CNPq--Brazil),
Instituto Nacional de Ciência e Tecnologia
de Fluidos Complexos (INCT-FCx), and the Research
Foundations of the States of Rio de Janeiro (FAPERJ) and
São Paulo (FAPESP). 

L.I. was supported by the project No. 2022/47/P/ST3/01236 co-funded by the National Science Centre and the European Union's Horizon 2020 research and innovation programme under the Marie Sk{\l}odowska-Curie grant agreement No. 945339.  Institutional and infrastructural support for the ENSEMBLE3 Centre of Excellence was provided through the ENSEMBLE3 project (MAB/2020/14) delivered within the Foundation for Polish Science International Research Agenda Programme and co-financed by the European Regional Development Fund and the Horizon 2020 Teaming for Excellence initiative (Grant Agreement No. 857543), as well as the Ministry of Education and Science initiative “Support for Centres of Excellence in Poland under Horizon 2020” (MEiN/2023/DIR/3797).

\bibliography{references.bib}

\end{document}